\documentclass[submission,copyright,creativecommons]{eptcs}
\providecommand{\event}{TFPIE 2024} 

\usepackage{amsmath}
\usepackage{tikz}
\usetikzlibrary{calc}
\usetikzlibrary{math}
\usetikzlibrary{arrows.meta}
\usepackage{iftex}

\newcommand{\vc}[1]{\vec{#1}}
\newcommand{\uvc}[1]{\hat{#1}}
\newcommand{\abs}[1]{\left| #1 \right|}

\newcommand{\yhat}{\uvc{y}}
\newcommand{\zhat}{\uvc{z}}
\newcommand{\phihat}{\hat{\phi}}
\newcommand{\thetahat}{\hat{\theta}}
\newcommand{\rhat}{\hat{r}}
\newcommand{\shat}{\hat{s}}

\newcommand{\del}{\vc{\nabla}}

\ifpdf
  \usepackage{underscore}         
  \usepackage[T1]{fontenc}        
\else
  \usepackage{breakurl}           
\fi

\title{Functional Programming in Learning Electromagnetic Theory}

\author{Scott N. Walck
\institute{Department of Physics\\
  Lebanon Valley College\\
  Annville, Pennsylvania, USA}
\email{walck@lvc.edu}
}

\def\titlerunning{Learning Electromagnetic Theory}
\def\authorrunning{Scott N. Walck}

\begin{document}

\maketitle

\begin{abstract}
Electromagnetic theory is central to physics.
An undergraduate major in physics typically takes a semester
or a year of electromagnetic theory as a junior or senior, and a graduate student
in physics typically takes an additional semester or year at a more advanced
level.
In fall 2023, the author taught his undergraduate electricity and magnetism
class using numerical methods in Haskell in parallel with traditional analytical methods.
This article describes what functional programming has to offer to physics
in general, and electromagnetic theory in particular.
We give examples from vector calculus, the mathematical language in which electromagnetic
theory is expressed, and electromagnetic theory itself.
\end{abstract}

\section{Introduction}

James Clerk Maxwell put the finishing touches on modern electromagnetic theory
in 1865, publishing the famous equations that still bear his name.
It is the oldest piece of theoretical physics that is part of humanity's current
best understanding of the physical world.

Electromagnetic (EM) theory describes one of the four fundamental forces of nature.
Today's physicists believe that all known interactions can be classified as
one of four forces.  These are the strong force, the electromagnetic force,
the weak force, and gravity.
Electricity holds atoms together, which seems enough to give it a claim to importance.

Electromagnetic theory serves as the model for modern field theories of elementary particles.
EM theory is the prototypical example of a \emph{field theory},
that is, a theory whose important quantities depend on space or spacetime.
Modern theories of elementary particle physics are based on
quantum field theories, the quantum versions of theories like
electromagnetic theory.

Surprisingly, Maxwell's 1865 electromagnetic theory did not need to be modified to
incorporate Einstein's 1905 relativity theory or the quantum theory of 1925.
In the case of relativity theory, electromagnetic waves began to be interpreted
differently, as physicists discarded the notion of an ether in which
EM waves traveled.
In the case of quantum theory,
mathematical objects like electric field and magnetic field
were interpreted differently, as operators rather than numbers or vectors,
but the Maxwell equations remained intact.
It seems that Faraday and Maxwell were onto something
fundamental about the universe.

Electromagnetic theory is the earliest theory that is still part of our current best understanding of the universe.
Newtonian mechanics is incredibly useful, and beautiful, but it cannot be
said to express our current best ideas about the universe.
The 20th-century ideas of quantum mechanics and relativity have each
led to newer theories that are slightly different from Newtonian mechanics,
although they give essentially the same answers for slow massive things.
Electromagnetic theory, on the other hand, passed unchanged through the
20th-century quantum and relativity revolutions.

Electromagnetic theory unites electricity, magnetism, and light into a single theory.
Light is an electromagnetic wave, a wave of electric and magnetic fields.
Electromagnetic theory is three theories in one, describing electricity, magnetism, and optics.

We take the attitude of Papert\cite{papert}
and others\cite{sicm,sussmanFDG,walck2014,walck2016}
that students are aided in their learning by having
building blocks with which to create interesting structures,
that such creative activity is a motivating and effective
way to learn, and that the feedback provided by
computer-language-based building blocks can expose our
confusions and produce delight in our achievements.

The author has written previously about the benefits of functional
programming to physics.  In \cite{walck2014}, we showed how
Newtonian mechanics
benefits from expression in a functional programming language.
In \cite{walck2016}, we showed how a course in quantum mechanics can benefit
from functional programming.
The textbook \cite{walck2023learn} treats both undergraduate Newtonian mechanics
and electromagnetic theory.  It can be used as a textbook for a course in
computational physics or as a companion text for courses in classical mechanics
or electromagnetism.  The author of this paper uses \cite{walck2023learn} in
both ways.  The present paper focuses on functional programming in electromagnetic
theory, and in particular on the author's experience incorporating it
into a traditional undergraduate electromagnetic theory course.

The plan for the paper is as follows.
In Section~\ref{sec:offer}, we describe what functional programming
has to offer to physics in general and to electromagnetic theory in particular.
Section~\ref{sec:field} describes field theory and how it can be encoded in
Haskell.
In Section~\ref{sec:vectorcalc}, we give examples of what vector calculus can
look like in a functional programming language.
Section~\ref{sec:biotsavart} describes an example of electromagnetic theory,
namely how magnetic field is produced by electric current.

\section{What Does Functional Programming Offer to Physics Pedagogy?}
\label{sec:offer}

In this section, we describe how programming in general, and functional programming
in particular, can help people learn physics, especially electromagnetic theory.
The first three items deal with programming in general, while the last five
focus on functional programming.

\subsection{With a little programming, we can study physical situations
  that are not exactly solvable.}
\label{sec:notexact}

Most physical situations are not exactly solvable.  The
traditional tools of algebra, calculus, and differential
equations only get us so far.  They can't solve most problems.
The theory tells you how to make (partial) differential
equations, but you can't solve them exactly.  We don't know how
to write down functions that exactly satisfy the differential
equations.  Traditionally, you spend a lot of time studying
situations that \emph{can} be solved exactly, and then come up
with tricks to approximate the other situations that can't be
solved exactly.

In EM theory, the electric field inside a capacitor with infinite
plates is exactly solvable, but the electric field inside a
capacitor with finite plates is not.  The magnetic field produced
by a circular loop carrying current is exactly solvable along the
axis that runs through the center of the circle (perpendicular to
the plane of the circle), but not exactly solvable anywhere else.

There are some simple, standard approximations (numerical
methods), that are known to be reasonably good, and don't involve
tricks.  The problem is that they require an enormous number of
exceedingly boring computations.  And here is where the computer
comes to the rescue.  Just by being able and willing to do an
enormous number of simple, boring computations, the computer can
give us good, approximate solutions to many problems.

\begin{figure}
\includegraphics[width=0.45\textwidth]{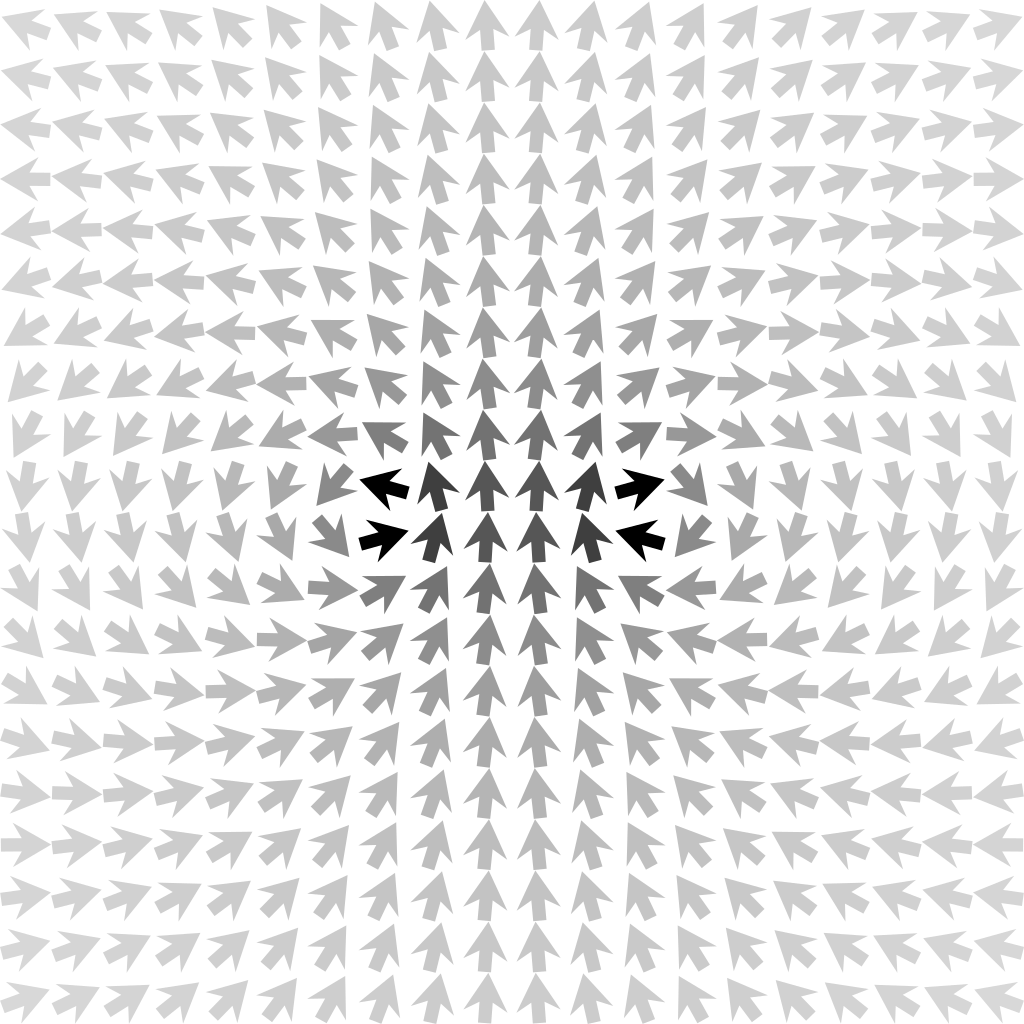}
\caption{Magnetic field produced by a current loop.  The current loop lies in the
$xy$ plane.  The figure shows the $yz$ plane.  A darker arrow indicates a stronger
magnetic field.
The current loop (not shown) pierces
the $yz$ plane near the darkest arrows.
This situation, simple as it seems, is exactly solvable only on the $z$ axis,
running vertically through the center of the figure.
}
\label{fig:bfielddipole}
\end{figure}

Figure \ref{fig:bfielddipole} shows the numerically calculated magnetic
field of a circular current loop, a problem which, surprisingly,
is not analytically solvable.

\subsection{Computer programming is a modern tool that can help solve problems.}

Why not use all the tools at our disposal, and in particular the
computer, to help us solve our problems?  Computers can do things
that calculators can't.  I don't predict the death of the pocket
calculator, but who knows?  The slide rule is, if not dead, an
eclectic tool used by very few people.

Physics societies like the American Association of Physics
Teachers (AAPT) have been encouraging physics teachers to include
more computer techniques in their courses.

\subsection{Programming is a valuable skill in itself.}

Programming, like mathematics and like writing, is a valuable skill in itself.  Knowing
something about programming is useful for getting a job, but it's
also useful for organizing your thinking.
(Mathematics and writing also help us organize our thinking.)
Writing code is not
just about getting the computer to do something.  It's also about
expressing ideas using formal language in a way that makes sense
to people.  And so, for the same reason that essay writing can
help you clarify your ideas about art or politics, code writing
can help you clarify your ideas about physics.

\subsection{Writing in a language that will be read by a computer forces a precision
  that human language and mathematical notation sometimes lack.}

In the preface to their wonderful book
\emph{Structure and Interpretation of Classical Mechanics} (SICM)\cite{sicm},
Sussman and Wisdom note
\begin{quote}
The traditional use of ambiguous notation is convenient in simple
situations, but in more complicated situations it can be a
serious handicap to clear reasoning. In order that the reasoning
be clear and unambiguous, we have adopted a more precise
mathematical notation. Our notation is functional and follows
that of modern mathematical presentations.
\end{quote}
and then
\begin{quote}
We require that our mathematical notations be explicit and
precise enough that they can be interpreted automatically, as by
a computer.
\end{quote}

Sussman was the first pioneer in the application of functional programming
to physics education.  He showed the power of higher-order functions
in physics using \emph{Scheme}, a dialect of Lisp he co-invented with
Steele, and used with Abelson in their classic computer science text
\emph{Structure and Interpretation of Computer Programs}\cite{sicp}.

The approach described in this paper is similar to SICM
in that both are interested in using functional programming
as a language for physics.
Our approach differs from SICM in several ways.
First, we use Haskell instead of Scheme, a difference which is noticeable
primarily in our reliance on types and Haskell's type system.
Second, SICM is a graduate-level textbook while we target undergraduates.
Third, SICM restricts its attention to classical mechanics, the theory of motion
created by Newton and developed by Lagrange and Hamilton, while our concern
in this paper is with a different theory of physics, namely electromagnetic theory.

\subsection{Avoiding mutation makes a programming language closer to mathematics.}

The core of a pure functional programming language,
while providing many interesting challenges for the compiler designer,
presents a very simple model of computation to the user of the language.
Names stand for values, not memory locations.
Names refer to a single thing, and don't change over time.

These are the principles upon which mathematical notation is based.
Physics already uses mathematical language, so it is a smaller step
for the physics student to learn the core of a pure functional programming
language than it is to learn an imperative language.  Obviously, there are
trade-offs; functional programming's affection for recursion
can make teaching iteration more difficult.

In my courses, I teach students how to use the Haskell functions
\texttt{map}, \texttt{iterate}, \texttt{filter}, \texttt{take}, and \texttt{takeWhile}
as well as list comprehensions,
but I do not teach them how to write explicitly
recursive functions.  Teaching recursion is a substantial endeavor,
and I need to use the time to teach ideas of physics.

\subsection{Types in a language help the reader understand and the writer
  clarify and organize.}
\label{sec:types}

I have come to believe that a strong system of types is exceedingly
helpful in a language for physics.
Algebraic data types are useful in a language for physics
for the same reason they are helpful in any domain-specific language,
namely that the key ideas of the domain can be encoded as types.

The code writer can experience the benefits of clarity and organization
of types even if she is not the person writing the data type definition.

An example of an algebraic data type in physics is \emph{charge distribution}.
Physicists like to talk about point charges, but they also like
to talk about electric charge that is distributed along a curve,
across a surface, or throughout a volume.
A \emph{charge distribution} is a specification of electric charge
that could be at a point, along a curve, 
across a surface, throughout a volume, or some combination of these.

We use an algebraic data type to make charge distribution into a type.
\begin{verbatim}
data ChargeDistribution
    = PointCharge   Charge      Position
    | LineCharge    ScalarField Curve
    | SurfaceCharge ScalarField Surface
    | VolumeCharge  ScalarField Volume
    | MultipleCharges [ChargeDistribution]
\end{verbatim}
The definition tells us that a charge distribution could be a point
charge, a line charge (physicists use the term \emph{line charge}
for any one-dimensional distribution of charge; it need not occur along
a straight line), a surface charge, a volume charge, or a combination
of these.
The \texttt{ChargeDistribution} data type is algebraic in having
multiple constructors, and recursive in the last constructor.

We can see from its definition that the
\texttt{ChargeDistribution} data type
is based on other data types for physics like
\texttt{Charge}, \texttt{Position}, \texttt{ScalarField},
\texttt{Curve}, \texttt{Surface}, and \texttt{Volume}.
\texttt{Charge} is just a type synonym for double-precision floating
point number.
\texttt{Position} is a point in three-dimensional space.
We'll discuss the \texttt{Position} data type in the next section.
\texttt{Curve}, \texttt{Surface}, and \texttt{Volume}
are geometric objects that describe finite curves, surfaces, and volumes.
These are helpful because EM theory has a substantial geometric content.

I do not demand that students be able to write algebraic data type definitions like
this.  Some students are interested in how this works, and I'm always
happy to share what I know, but my main interest is in having students
use this data type to describe a charge distribution.
Using this data type brings one face to face with precisely what a charge distribution
is, how we talk about it, and what's it's good for.

In EM theory, scalar and vector fields play a prominent role.
A \emph{field} in physics is a function of space or spacetime.
In an undergraduate setting, it is more common to treat fields as
functions of space, and I follow that custom with the
following type synonyms for scalar field and vector field.
\begin{verbatim}
type ScalarField = Position -> R
type VectorField = Position -> Vec
\end{verbatim}
In the first definition, \texttt{R} is a type synonym for \texttt{Double},
a double-precision floating point number.  I like to think of these numbers
as approximations to real numbers, and they occur so often in physics
that it is convenient to give them a shorter name that comes closer to how
we think about them.
In the second definition, \texttt{Vec} is a data type for three-dimensional
vectors.  Such vectors are ubiquitous in physics, especially at the undergraduate
pre-relativity level.

The central idea of electrostatics, the portion of EM theory focusing
on electrical phenomena when nothing is changing in time,
is that a charge distribution produces an electric field.
We encode this central idea as a function.
\begin{verbatim}
eField :: ChargeDistribution -> VectorField
\end{verbatim}
The type signature of \texttt{eField} sings the central idea of electrostatics.
The definition of \texttt{eField} takes some effort to develop,
as we need various kinds of integrals in order to deal with the
continuous charge distributions.
The type signature, however, serves as a wonderful summary and reminder
of the big picture, and hopefully serves as a counterbalance to the possibility
that people might get lost in the details of the function definition.

In fact, students routinely get lost in the traditional analytical
presentation of electric field produced by a line charge, electric field
produced by a surface charge, and electric field produced by a volume charge.
There are so many details to worry about, and nothing in the traditional
mathematical notation is reminding us of the central idea.
A language with a strong system of types can help us keep the important
ideas in mind.

\subsection{Higher-order functions are a natural way to express the higher-order ideas
  that physics trades in.}

Take the derivative from calculus as a higher-order idea that physics uses.
The derivative of a function at a point expresses the rate at which the function
changes at that point.
We can write a numerical derivative in Haskell.

\begin{verbatim}
type R = Double

derivative :: R -> (R -> R) -> R -> R
derivative dt f t = (f (t + dt/2) - f (t - dt/2)) / dt
\end{verbatim}
In this definition, \texttt{dt} is a numerical step size for the numerical derivative
we are calculating.  The second argument is the function \texttt{f :: R -> R}
we want to differentiate.  The third argument \texttt{t :: R} is the value
of the independent variable where we want to estimate the slope.

Physicists like to think of the derivative as an ``operator'' that takes a function
as input and gives another function as output.
Haskell's currying allows us to think just like that.
Instead of thinking of the \texttt{derivative} as a function that takes
three inputs (step size, function, and independent variable), we can think
of \texttt{derivative} as a function that takes two inputs
(step size and function) and returns a function.
Even better, we can come to see that these two ways of thinking
are the same for the computer because there is a sense in which
there is no difference.  The two ways of thinking are equivalent.
Better yet, the function \texttt{derivative 1e-6} really is
the ``operator'' that takes a function as input and gives a function as output.

Numeric integration is another example of a higher-order function
that encodes a basic idea of physics.  And since calculus,
vector calculus, and differential equations are so ubiquitous in physics,
there are a plethora of higher-order functions serving calculus-related
roles, some of which we will see later in this article.

Electromagnetic theory is primarily about vector
fields, namely the electric and magnetic fields.
We saw above that a vector field is a function, so
any function that deals with a vector field is a higher-order
function.  

In physics, ideas we regard as basic, like the derivative,
the integral, and the vector field, naturally invite us to
use higher-order functions.
Because higher-order functions in Haskell and other functional programming
languages are easy to write and use,
the language of the code can help you understand the theory.
When a language allows a succinct expression of an idea that a domain
regards as basic, using that language
allows one to experiment and play with the idea,
which is a great way to learn.
Reading and writing in a language that makes it easy (or at least easier)
to express the ideas in a theory promotes understanding of the theory.
I claim that typed functional languages are especially suitable for physics,
and I show examples below as evidence of this claim.
I have anecdotal evidence that students understand analytical methods
in electromagnetic theory better for having studied numerical methods
in a functional programming language.

\subsection{Functional programming offers a vision of a language for both calculating
  and proving, two activities that theoretical physics is full of.}

Physics has some results that can be obtained from deductive reasoning.
The Maxwell equations imply conservation of electric charge, for example.
In a course in EM theory, we show how to derive the continuity equation,
an expression of local charge conservation, from the Maxwell equations.

This derivation takes a handful of steps, and although it doesn't
rise to the stature of a theorem as appreciated by mathematicians,
nevertheless the language and pattern of a theorem are present.
If the Maxwell equations are true, then the continuity equation is
true.  Such relationships are helpful to recognize, because a deep
understanding of any theory requires going beyond ``what is true''
into knowing when ideas are independent, and knowing when one
idea is a logical consequence of another.

Haskell, the language that I use in my EM theory class,
is great for calculation, but not really intended for theorem proving.
However, a handful of functional languages based on dependent types,
such as Coq, Agda, Idris, and Lean, are capable of theorem proving
(meaning proof checking with some automation)
in addition to calculation.

It seems that a language that could calculate and prove would
be a wonderful language in which to express the principles
of physics.  While I have great interest in this possibility,
it is in no sense a reality in my teaching.

It seems to me that the possibility of using a dependently-typed
functional language for physics in the future
is a reason to learn a functional language like Haskell for physics now.

\section{Field Theory in Haskell}
\label{sec:field}

Electromagnetic theory is a \emph{field theory}, where a \emph{field}
in physics is a quantity that depends on location in space or spacetime.
(The definition of \emph{field} in mathematics is different, and unrelated.)
The most common quantities that depend on space or spacetime are
scalars (basically numbers) and vectors.
In undergraduate electromagnetic theory, it is common to view fields
as functions of space, and we will follow this practice.
A scalar quantity that depends on space is called a \emph{scalar field},
and a vector quantity that depends on space is called a \emph{vector field}.
We gave type definitions for \texttt{ScalarField} and \texttt{VectorField}
in section \ref{sec:types}.

EM theory can be expressed (quite beautifully, in fact) in relativistic notation
where the vectors are 4-vectors
(elements of a 4-dimensional vector space)
and the fields are functions of spacetime,
but in an undergraduate setting, it is much more common to use a notation in
which vectors are 3-vectors (members of a 3-dimensional vector space),
and the fields are functions of space.  This does not preclude the possibility
that fields can change in time; it merely places space and time on an
apparently different footing.

\begin{figure}
  \tikzset{
    MyPersp/.style={x={(1,0)},y={(0,1)},z={(-0.4,-0.4)}}
  }
  \begin{tikzpicture}[MyPersp,>=Stealth]
    \tikzmath{
      \radius=5;
      \angletheta=50;
      \anglephi=45;
      \costheta=cos(\angletheta);
      \sintheta=sin(\angletheta);
      \cosphi=cos(\anglephi);
      \sinphi=sin(\anglephi);
      \x=\radius*sin(\angletheta)*cos(\anglephi);
      \y=\radius*sin(\angletheta)*sin(\anglephi);
      \z=\radius*cos(\angletheta);
      \s=\radius*sin(\angletheta);
      \unitvec=1;
      \arctheta=1.5;
      \arcphi=1;
    }
    \draw (0,0,0) -- (5,0,0);
    \draw (0,0,0) -- (0,5,0);
    \draw (0,0,0) -- (0,0,5);

    \draw (5,0,0) node[below] {$y$};
    \draw (0,5,0) node[left] {$z$};
    \draw (0,0,5) node[left] {$x$};

    \draw[fill] (\y,\z,\x) circle (0.05);

    \draw ( 0,    \z,0) -- (\y,\z,\x);
    \draw (0.5*\y,\z,0.5*\x) node[above] {$s$};

    \draw (0,0,\arcphi) \foreach \phi in {5,10,...,\anglephi} {--({\arcphi*sin(\phi)},0,{\arcphi*cos(\phi)})};
    \draw ({\arcphi*sin(0.5*\anglephi)},0,{\arcphi*cos(0.5*\anglephi)}) node[below] {$\phi$};

    \draw (\y, 0,\x) -- (\y,\z,\x);
    \draw (\y,0.5*\z,\x) node[right] {$z$};

    \draw ( 0, 0, 0) -- (\y, 0,\x);
    \draw[dashed] (\y, 0,\x) -- (\y, 0, 0);
    \draw[dashed] (\y, 0,\x) -- ( 0, 0,\x);

    \draw[thick,->] (\y,\z,\x) -- +(                0 ,\unitvec,                 0 );
    \draw[thick,->] (\y,\z,\x) -- +({\unitvec*\sinphi}, 0      ,{ \unitvec*\cosphi});
    \draw[thick,->] (\y,\z,\x) -- +({\unitvec*\cosphi}, 0      ,{-\unitvec*\sinphi});

    \draw ($(\y,\z,\x)+(                0 ,\unitvec,                 0 )$) node[right] {$\zhat$};
    \draw ($(\y,\z,\x)+({\unitvec*\sinphi}, 0      ,{ \unitvec*\cosphi})$) node[right] {$\shat$};
    \draw ($(\y,\z,\x)+({\unitvec*\cosphi}, 0      ,{-\unitvec*\sinphi})$) node[right] {$\phihat$};
\end{tikzpicture}
  \begin{tikzpicture}[MyPersp,>=Stealth]
    \tikzmath{
      \radius=5;
      \angletheta=50;
      \anglephi=45;
      \costheta=cos(\angletheta);
      \sintheta=sin(\angletheta);
      \cosphi=cos(\anglephi);
      \sinphi=sin(\anglephi);
      \x=\radius*sin(\angletheta)*cos(\anglephi);
      \y=\radius*sin(\angletheta)*sin(\anglephi);
      \z=\radius*cos(\angletheta);
      \s=\radius*sin(\angletheta);
      \unitvec=1;
      \arctheta=1.5;
      \arcphi=1;
    }
    \draw (0,0,0) -- (5,0,0);
    \draw (0,0,0) -- (0,5,0);
    \draw (0,0,0) -- (0,0,5);

    \draw (5,0,0) node[below] {$y$};
    \draw (0,5,0) node[left] {$z$};
    \draw (0,0,5) node[left] {$x$};

    \draw[fill] (\y,\z,\x) circle (0.05);

    \draw ( 0, 0, 0) -- (\y,\z,\x);
    \draw (0.7*\y,0.7*\z,0.7*\x) node[above] {$r$};

    \draw (0,\arctheta,0) \foreach \theta in {5,10,...,\angletheta}
          {--({\arctheta*sin(\theta)*\sinphi},{ \arctheta*cos(\theta)},{ \arctheta*sin(\theta)*\cosphi})};
    \draw ({\arctheta*sin(0.5*\angletheta)*\sinphi},{ \arctheta*cos(0.5*\angletheta)},{ \arctheta*sin(0.5*\angletheta)*\cosphi})
          node[right] {$\theta$};

    \draw (0,0,\arcphi) \foreach \phi in {5,10,...,\anglephi} {--({\arcphi*sin(\phi)},0,{\arcphi*cos(\phi)})};
    \draw ({\arcphi*sin(0.5*\anglephi)},0,{\arcphi*cos(0.5*\anglephi)}) node[below] {$\phi$};

    \draw (\y, 0,\x) -- (\y,\z,\x);
    \draw ( 0, 0, 0) -- (\y, 0,\x);
    \draw[dashed] (\y, 0,\x) -- (\y, 0, 0);
    \draw[dashed] (\y, 0,\x) -- ( 0, 0,\x);

    \draw[thick,->] (\y,\z,\x) -- +({\unitvec*\sintheta*\sinphi},{ \unitvec*\costheta},{ \unitvec*\sintheta*\cosphi});
    \draw[thick,->] (\y,\z,\x) -- +({\unitvec*\costheta*\sinphi},{-\unitvec*\sintheta},{ \unitvec*\costheta*\cosphi});
    \draw[thick,->] (\y,\z,\x) -- +({\unitvec          *\cosphi},                   0 ,{-\unitvec          *\sinphi});

    \draw ($(\y,\z,\x)+({\unitvec*\sintheta*\sinphi},{ \unitvec*\costheta},{ \unitvec*\sintheta*\cosphi})$) node[above] {$\rhat$};
    \draw ($(\y,\z,\x)+({\unitvec*\costheta*\sinphi},{-\unitvec*\sintheta},{ \unitvec*\costheta*\cosphi})$) node[right] {$\thetahat$};
    \draw ($(\y,\z,\x)+({\unitvec          *\cosphi},                   0 ,{-\unitvec          *\sinphi})$) node[right] {$\phihat$};
\end{tikzpicture}
\caption{Cylindrical coordinates (left) and spherical coordinates (right)}
\label{coordinatesystems}
\end{figure}
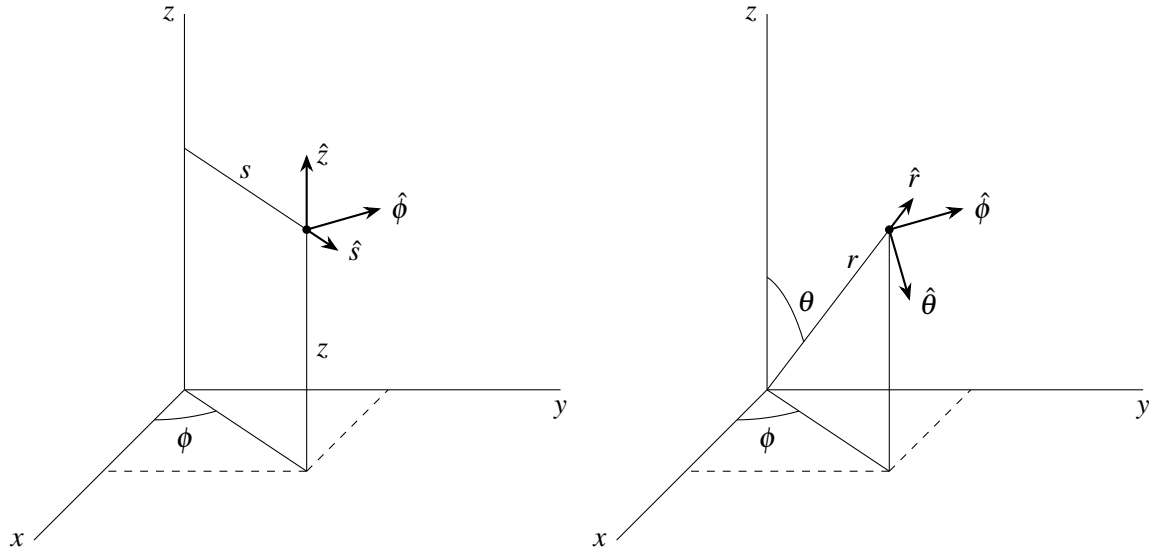

Having decided to pursue an undergraduate notation in which
3-dimensional space is central, we turn to the question of how
to represent points in 3-dimensional space.
Our first choice is Cartesian coordinates $x, y, z$, in which the
three coordinate axes are mutually perpendicular.
However, several important situations in EM theory have spherical or cylindrical symmetry,
which motivates the use of spherical and cylindrical coordinates.
Even situations which do not have full cylindrical or spherical symmetry
can benefit from these coordinates.
In fact, there are many possible coordinate systems for 3-dimensional space,
but we will focus on the 3 most commonly used, namely Cartesian, cylindrical,
and spherical.  Figure~\ref{coordinatesystems} shows how cylindrical and spherical
coordinates are defined.

In cylindrical coordinates, we use $s$ to represent the distance from the $z$~axis,
$\phi$ to represent the angle shown in Figure~\ref{coordinatesystems}, and $z$ to
mean the same thing it means in Cartesian coordinates.  The cylindrical coordinates
of a point in space are then given by the triple $(s,\phi,z)$.
In spherical coordinates, we use $r$ to represent the distance from the origin,
$\theta$ to represent the angle shown in Figure~\ref{coordinatesystems}, and $\phi$ to
mean the same thing it means in cylindrical coordinates.  The spherical coordinates
of a point in space are then given by the triple $(r,\theta,\phi)$.
This notation for cylindrical and spherical coordinates comes from Griffiths'
popular textbook\cite{griffiths2017introduction}.

The purpose of the \texttt{Position} data type is to free our minds
from the detail of establishing a single, canonical way of referring
to a point in space.  We wish to freely use any of the three coordinate
systems in constructing a \texttt{Position}, and any of the three
in extracting coordinates.  Here is the definition of the \texttt{Position}
data type.

\begin{verbatim}
data Position = Cart R R R
                deriving (Show)
\end{verbatim}

We store coordinates in Cartesian form,
and name the constructor \texttt{Cart} to remind us of this fact,
but users of \texttt{Position} need not be aware of the specifics
of how the coordinates are stored.

The following three functions construct
a \texttt{Position} from coordinates.

\begin{verbatim}
cartesian   :: (R,R,R) -> Position   -- (x,y,z)
cylindrical :: (R,R,R) -> Position   -- (s,phi,z)
spherical   :: (R,R,R) -> Position   -- (r,theta,phi)
\end{verbatim}
For convenience, we define curried forms \texttt{cart}, \texttt{cyl}, and \texttt{sph}.

Finally, we have three functions to extract coordinates from a \texttt{Position}
in any of the three systems.
\begin{verbatim}
cartesianCoordinates   :: Position -> (R,R,R)   -- (x,y,z)      
cylindricalCoordinates :: Position -> (R,R,R)   -- (s,phi,z)    
sphericalCoordinates   :: Position -> (R,R,R)   -- (r,theta,phi)
\end{verbatim}

Since the \texttt{VectorField} is the type of the electric field and the magnetic field,
and since we can develop some geometric intuition by making pictures of vectors fields,
we spend some time doing this in my class.  The function \texttt{vfGrad} uses a technique I call
\emph{gradient visualization}, in which the strength of the field is denoted by
the darkness of arrows.  This is the kind of visualization used in Figure~\ref{fig:bfielddipole}
for the magnetic field produced by a circular current loop.
\begin{verbatim}
vfGrad :: (R -> R)
       -> ((R,R) -> Position)
       -> (Vec -> (R,R))
       -> FilePath
       -> Int    -- n for n x n
       -> VectorField
       -> IO ()
\end{verbatim}
There are several parameters that the function
\texttt{vfGrad} needs in order to produce a picture like
Figure~\ref{fig:bfielddipole}.
The first argument to \texttt{vfGrad} is a monotonic function
that allows the transition from light arrows to dark arrows
to occur in a nonlinear way; using this we can avoid pictures
with a small number of black arrows on an otherwise white background
that would occur with a linear scaling because the field is very
strong at one location.
The next two arguments to \texttt{vfGrad} control the relationship
between the 3-dimensional field and the 2-dimensional picture;
they specify which plane we are looking at and how to convert
3-dimensional vectors into 2-dimensional arrows.
The last three arguments to \texttt{vfGrad} are a file name
for the graphics file, an integer describing how many arrows
we want in each direction, and finally the vector field itself.
The definition of this function uses Brent Yorgey's
\emph{diagrams} package\cite{diagrams}.

With types for scalar fields and vector fields, we are ready
to explore the mathematics used in electromagnetic theory,
vector calculus.

\section{Vector Calculus in Haskell}
\label{sec:vectorcalc}

Vector calculus is a large subject.
Undergraduate physics majors at Lebanon Valley College take one
semester of vector calculus, but that is not quite enough
for them to get to results like Stokes' theorem, which we need
in electromagnetic theory.
So, part of our electricity and magnetism course is a review
and deeper dive into vector calculus.

I have always included some vector calculus in my electricity and
magnetism course, but in fall 2023,
I taught analytical methods for vector calculus in parallel with
numerical methods using Haskell.
This section gives a sampling of what vector calculus can
look like in Haskell.

In order to do vector calculus, we are first going to need a
data type for vectors.  The \texttt{Vec} type was seen in passing
in Section~\ref{sec:types}, and is for 3-dimensional vectors.
Details about the \texttt{Vec} type can be found in
\cite{walck2014} and \cite{walck2023learn}.

Electromagnetic theory is a geometric subject,
and the vector calculus it uses has a geometric flavor.
To see why, consider the following form of the Maxwell equations.

\subsection{Maxwell Equations, Integral Form}

There are several ways of writing the Maxwell equations.
Below we give a version known as the ``integral form''.
(There is also a ``differential form'' that is probably
seen more often, but hides the geometric character of the
theory.)
For each of the four equations, I give a statement in words,
followed by an equation in mathematical notation that expresses
the same thing.

\begin{enumerate}
\item \textbf{Ampere-Maxwell Law:}
The rate of change of electric flux through a surface
is the magnetic circulation around its boundary
minus the current flowing through the surface.
\[
\frac{d}{dt} \int_S \vc{E} \cdot d\vc{a}
  = \frac{1}{\epsilon_0 \mu_0} \int_{\partial S} \vc{B} \cdot d\vc{\ell}
    - \frac{1}{\epsilon_0} I
\]
\item \textbf{Faraday's Law:}
The rate of change of magnetic flux through a surface
is the opposite of the electric circulation around its boundary.
\[
\frac{d}{dt} \int_S \vc{B} \cdot d\vc{a}
  = - \int_{\partial S} \vc{E} \cdot d\vc{\ell}
\]
\item \textbf{Gauss's Law:}
The electric flux through a closed surface is
the charge contained inside.
\[
\int_S \vc{E} \cdot d\vc{a}
  = \frac{1}{\epsilon_0} Q
\]
\item \textbf{No Magnetic Monopoles Law:}
The magnetic flux through a closed surface is zero.
\[
\int_S \vc{B} \cdot d\vc{a} = 0
\]
\end{enumerate}

In these equations, $\vc{E}$ is the electric field and
$\vc{B}$ is the magnetic field.
Electric and magnetic fields are the fundamental physical quantities
that electromagnetic theory deals with.  They are created by charged particles,
and they exert forces on charged particles.
Most (perhaps all, I have no counterexample) physicists regard electric and
magnetic fields to be just as real as the particles they interact with.
Particles and fields, then, have a symbiotic relationship.
Conceptually, they are equal partners in the explanation of electromagnetic
phenomena, both dynamic actors on the electromagnetic stage.
Most people, students included, have a much easier time imagining a particle
than imagining a field.  Electric and magnetic fields have type
\texttt{VectorField}, which is a type synonym for \texttt{Position -> Vec}.

In the first two equations, $S$ is a surface and $\partial S$
is its boundary curve.  An electric current $I$ flows through surface $S$.
In the third and fourth equations, $S$ is a closed surface,
containing some volume, and $Q$ is the total electric charge
in that volume.
The constants $\epsilon_0$ and $\mu_0$ appear in the equations because
they use the International System (SI) of units.
These constants are elided from the English-language descriptions.

Notice that surfaces and their boundary curves make an explicit appearance
in this form of the Maxwell equations.
This is evidence of the geometrical character
of electromagnetic theory.

Beyond their role in the statement of the Maxwell equations, curves and
surfaces are used to describe continuous charge and current distributions.
We saw an example of this with the definition of the \texttt{ChargeDistribution}
type in Section~\ref{sec:types}.  The \texttt{ChargeDistribution} data type makes reference
to the types \texttt{Curve}, \texttt{Surface}, and \texttt{Volume}.
The simplest of these is the \texttt{Curve} data type, defined as follows.
\begin{verbatim}
data Curve = Curve { curveFunc          :: R -> Position
                   , startingCurveParam :: R  -- t_a
                   , endingCurveParam   :: R  -- t_b
                   }
\end{verbatim}
A \texttt{Curve} is a map of one parameter into space, along with initial and final values
of that parameter.
A \texttt{Surface} is a map of two parameters into space, along with boundary data.
\begin{verbatim}
data Surface = Surface ((R, R) -> Position) R R (R -> R) (R -> R)
\end{verbatim}
The first argument to \texttt{Surface} is
the parameterizing function that maps $(s, t)$
into a \texttt{Position}.  The second and third arguments are lower and upper
limits of the parameter $s$.
The fourth and fifth arguments are functions of $s$ that give the lower
and upper limits of $t$.  Allowing the limits of the second parameter
to depend on the first parameter makes it easier to describe a
surface like a triangle.
The boundary of a \texttt{Surface} is a \texttt{Curve}.

A \texttt{Volume} is a map of three parameters into space, along with boundary information.
The boundary of a \texttt{Volume} is a \texttt{Surface}.

Once we have data types for curves, surfaces, and volumes,
vector calculus wants us to perform integrals over them.

\subsection{Flux and Circulation Integrals}
\label{sec:integrals}

For any vector field $\vc{F}$
and any surface $S$, the \emph{flux} of the field through the surface
is the integral
\[
\int_S \vc{F} \cdot d\vc{a} .
\]
The integral is calculated by approximating the surface $S$ by many small
triangles.  Each triangle has a vector area whose magnitude is the area
of the triangle and whose direction is perpendicular to the triangle.
At each triangle, we form the inner (dot) product of the vector field
$\vc{F}$ at the location of the triangle and the vector area of the triangle;
this inner product is a number.  We add up these numbers for all of the
triangles and call the result the flux of $\vc{F}$ through $S$.
If the vector field is the electric field, we call the flux electric flux.
If the vector field is the magnetic field, we call the flux magnetic flux.
If $\vc{F}$ represented the flow of stuff, then the flux of $\vc{F}$ through
$S$ is how much stuff flows through $S$.

Here is the Haskell code to approximate a flux integral.
\begin{verbatim}
dottedSurfaceIntegral :: SurfaceApprox -> VectorField -> Surface -> R
dottedSurfaceIntegral approx vF s
    = sum [vF r' <.> da' | (r',da') <- approx s]
\end{verbatim}
The function \texttt{dottedSurfaceIntegral} needs an approximation method
(type \texttt{SurfaceApprox})
to divide the surface into triangles.
\begin{verbatim}
type SurfaceApprox = Surface -> [(Position, Vec)]
\end{verbatim}
This approximation method,
called \texttt{approx} in the code, produces
a list of pairs of positions and vector areas.
The left side of the list comprehension shows that we take the dot product
(using the operator \texttt{<.>}) of the vector field \texttt{vF} evaluated
at position \texttt{r'} with the vector area \texttt{da'}.
The numerical results of the inner products are then added up with \texttt{sum}.

For any vector field $\vc{F}$
and any curve $C$, the \emph{circulation} of the field along the curve
is the integral
\[
\int_C \vc{F} \cdot d\vc{\ell} .
\]
The integral is calculated by approximating the surface $C$ by many small
segments.  Each segment has a vector length whose magnitude is the length
of the segment and whose direction is along the segment.
At each segment, we form the inner (dot) product of the vector field
$\vc{F}$ at the location of the segment and the vector length of the segment;
this inner product is a number.  We add up these numbers for all of the
segments and call the result the circulation of $\vc{F}$ along $C$.
If the vector field is the electric field, we call the circulation electric circulation.
If the vector field is the magnetic field, we call the circulation magnetic circulation.

Here is the Haskell code to approximate a circulation integral.
\begin{verbatim}
dottedLineIntegral :: CurveApprox -> VectorField -> Curve -> R
dottedLineIntegral approx f c = sum [f r' <.> dl' | (r',dl') <- approx c]
\end{verbatim}
The function \texttt{dottedLineIntegral} needs an approximation method
to divide the curve into segments.  This approximation method,
called \texttt{approx} in the code, produces
a list of pairs of positions and vector lengths.
The left side of the list comprehension shows that we take the dot product
(using the operator \texttt{<.>}) of the vector field \texttt{f} evaluated
at position \texttt{r'} with the vector length \texttt{dl'}.
The numerical results of the inner products are then added up with \texttt{sum}.

There are quite a number of these line, surface, and volume integrals
that are used in electromagnetic theory.  The table below shows
the nine integrals that EM theory needs from vector calculus.

\begin{center}
\begin{tabular}{cccc}
Haskell function               & mathematical notation      & \multicolumn{2}{c}{requires} \\ \hline
\texttt{scalarLineIntegral}    & $\int_C f \ d\ell$         & scalar field & curve \\
\texttt{vectorLineIntegral}    & $\int_C \vc{F} \ d\ell$    & vector field & curve \\
\texttt{dottedLineIntegral}    & $\int_C \vc{F} \cdot d\vc{\ell}$  & vector field & curve \\
\texttt{crossedLineIntegral}   & $\int_C \vc{F} \times d\vc{\ell}$ & vector field & curve \\
\texttt{scalarSurfaceIntegral} & $\int_S f \ d a$                  & scalar field & surface \\
\texttt{vectorSurfaceIntegral} & $\int_S \vc{F} \ d a$             & vector field & surface \\
\texttt{dottedSurfaceIntegral} & $\int_S \vc{F} \cdot d\vc{a}$     & vector field & surface \\
\texttt{scalarVolumeIntegral}  & $\int_V f \ d v$                  & scalar field & volume \\
\texttt{vectorVolumeIntegral}  & $\int_V \vc{F} \ d v$             & vector field & volume \\
\end{tabular}
\end{center}

Vector calculus has its own versions of the fundamental theorem of calculus,
the theorem that makes precise the idea that derivatives and integrals are
inverse operations of each other.
The fundamental theorems of vector calculus are known as the gradient theorem,
Stokes' theorem, and the divergence theorem.
They can be expressed using the integrals in the table above.

\subsection{A Homework Problem}

After students understand the meaning of the integrals in the table above,
they are in a position to study the three fundamental theorems of vector
calculus.  Since each theorem claims that an integral of a derivative
over some geometric object is equal to a different integral over the boundary of that
object, checking these fundamental theorems provides opportunities
to practice evaluating the integrals.  Even better, the two sides, which are
each a bit of a challenge to set up, must give the same number in the end,
which serves as satisfying evidence that the integrals were done correctly.

Stokes' theorem claims that for any vector field $\vc{F}$
and any surface $S$, the following equality holds.
\[
\int_S (\del \times \vc{F}) \cdot d\vc{a} = \int_{\partial S} \vc{F} \cdot d\vc{\ell}
\]
The expression $\del \times \vc{F}$ is called the curl of $\vc{F}$.
The curl is a vector derivative that takes a \texttt{VectorField} as input
and produces a \texttt{VectorField} as output.  The physical meaning of
the curl is the extent to which a vector field, if it were a velocity field of water,
encourages rotation of a small object placed in the water.
A vector field with zero curl is called conservative, meaning that a dotted line
integral of the field around any closed curve is zero.

The following homework problem, which was assigned in fall 2023,
asks students to check Stokes' theorem for a particular vector field
and a particular surface.
\begin{quote}
\textbf{Homework Problem: }
Check Stokes' theorem using
the vector field
\[
\vc{F}(x,y,z) = -z \yhat + y \zhat
\]
and
the rectangular region with corners
$(x,y,z) = (0,0,-4)$, $(0,2,-4)$, $(0,2,4)$, $(0,0,4)$,
in which the orientation is in the positive $x$ direction.

Checking the fundamental theorem for curls means (a) finding the curl
of your vector field, (b) finding the flux integral of the curl over the
surface, (c) finding the line integral (the circulation) of your vector
field over the boundary of the surface (which has 4 parts), and (d)
confirming that the results from (b) and (c) are the same. You may do
this problem analytically, numerically, or a combination of both.
\end{quote}

To solve this homework problem numerically with Haskell, we first encode the vector field
$\vc{F}$ as \texttt{vF} and the rectangle as \texttt{rect}.
\begin{verbatim}
vF :: VectorField
vF r = let (x,y,z) = cartesianCoordinates r
       in (-z) *^ yHat r ^+^ y *^ zHat r

rect :: Surface
rect = Surface (\(y,z) -> cart 0 y z) 0 2 (const (-4)) (const 4)
\end{verbatim}
To calculate the left-hand side of Stokes' theorem, we need to take the curl
of the vector field.  As with a numerical derivative, we need to give a step
size over which to evaluate the curl; we choose $10^{-6}$.
\begin{verbatim}
curlF :: VectorField
curlF = curl 1e-6 vF
\end{verbatim}
Now we can calculate the left side of Stokes's theorem.
\begin{verbatim}
leftSide :: R
leftSide = dottedSurfaceIntegral (surfaceSample 200) curlF rect
\end{verbatim}
We get the result \texttt{32.00000000236293}.
For the right side, we need the boundary curve of the reactangle.
\begin{verbatim}
boundaryParam :: R -> Position
boundaryParam t
    | t < 1      = cart 0 (2*t) (-4)
    | t < 2      = cart 0 2 (8*(t-1) - 4)
    | t < 3      = cart 0 (2 - 2 * (t-2)) 4
    | otherwise  = cart 0 0 (4 - 8 * (t-3))

boundaryOfRect :: Curve
boundaryOfRect = Curve boundaryParam 0 4

rightSide :: R
rightSide = dottedLineIntegral (curveSample 1000) vF boundaryOfRect
\end{verbatim}
The right side gives \texttt{32.0}, which is reasonably close to the left side,
and can be explained by numerical error.

In this homework problem, I give students the choice about whether to use
analytic (pencil and paper) methods, numerical methods, or a combination of both.
I do not always give students this choice.  Sometimes I demand they use
analytic methods; sometimes I demand they use numerical methods.
At least one third of the time, however, I give students the choice,
which they seem to appreciate.

For this homework problem in fall 2023, 5 of the 10 students who submitted it
chose to do it analytically, and 5 chose to do it numerically using Haskell.

As a last example of the power of types and higher order functions
in EM theory, we turn to calculation of the magnetic field by
the Biot-Savart law.

\section{Magnetic Field Produced by a Current Distribution}
\label{sec:biotsavart}

Earlier, we saw that the central idea of electrostatics
is that a charge distribution produces an electric field.
Analogously, the central idea of magnetostatics is that
a current distribution produces a magnetic field.
One can define a type \texttt{CurrentDistribution},
and a function
\begin{verbatim}
bField :: CurrentDistribution -> VectorField
\end{verbatim}
that calculates the magnetic field produced by any current
distribution.\cite{walck2023learn}

In this section, we focus on a particular type of current distribution,
the one we probably have in mind when thinking of electric current,
namely the current flowing through a wire.
The equation for calculating the magnetic field produced by current
flowing through a wire is called the Biot-Savart law, and is shown below.
\begin{equation}
\vc{B}(\vc{r}) = -\frac{\mu_0 I}{4 \pi} \int_C \frac{\vc{r} - \vc{r}'}{~\abs{\vc{r} - \vc{r}'}^3} \times d\vc{\ell}'
\label{eqnbiotsavart}
\end{equation}
In this equation, $C$ is a curve of arbitrary shape that describes the
configuration of the wire.  The wire could run along a straight line,
around in a circle, in a helix, or any other configuration.
The integral asks the \emph{source point} $\vc{r}'$ to take all possible values
along the wire, the current at each position along the wire being a source of magnetic field.
The \emph{field point} $\vc{r}$ is the place where we want to find the
magnetic field; it is a constant from the perspective of the integral.
The current $I$ is just a number, usually measured in Amperes.
The symbol $d\vc{\ell}'$ represents a small section of the wire.
The multiplication $\times$ is the vector cross product.

Here is the Biot-Savart law, translated into Haskell.
\begin{verbatim}
bFieldFromLineCurrent
    :: Current      -- current (in Amps)
    -> Curve
    -> VectorField  -- magnetic field (in Tesla)
bFieldFromLineCurrent i c r
    = let coeff = -mu0 * i / (4 * pi)  -- SI units
          integrand r' = d ^/ magnitude d ** 3
              where d = displacement r' r
      in coeff *^ crossedLineIntegral (curveSample 1000) integrand c
\end{verbatim}

The type \texttt{Current} is a synonym for \texttt{R} or \texttt{Double};
it is the number of Amperes of current flowing through the wire.
We define a local constant \texttt{coeff} to hold the numerical value
of $-\mu_0 I/4 \pi$ in SI units, and we define a local function \texttt{integrand}
to hold the integrand. 
We want to define a local variable
\texttt{d} for the displacement from \texttt{r'} to \texttt{r},
but because \texttt{r'} exists locally to the function \texttt{integrand},
the definition for \texttt{d} must occur within the definition
for \texttt{integrand} and cannot be placed parallel to the definitions
of \texttt{coeff} and \texttt{integrand}.
We use the \texttt{crossedLineIntegral} to do the integration;
this function appears in the table of vector calculus integrals
in Section~\ref{sec:integrals}.

The type signature of \texttt{bFieldFromLineCurrent}
makes clear, in a computer-checked way, the two inputs
required to find the magnetic field:
the \texttt{Current} and the \texttt{Curve} along which the current flows.
For a reader of Haskell, the function \texttt{bFieldFromLineCurrent}
is a clearer description of what is going on than Equation~\ref{eqnbiotsavart}
since the latter does not make it terribly clear that the magnetic field
depends only on the curve and the current.

The function \texttt{bFieldFromLineCurrent}
was used to produce Figure~\ref{fig:bfielddipole} in Section~\ref{sec:notexact}.

Next we describe a homework problem which requires the Biot-Savart law.
In Section \ref{sec:notexact}, we mentioned that
the magnetic field produced
by a circular loop carrying current is exactly solvable along the
axis that runs through the center of the circle (perpendicular to
the plane of the circle), but not exactly solvable anywhere else.
This homework problem asks students to use analytical methods
to find the magnetic field on the axis, where a closed-form
algebraic result can be obtained, and then to
use numerical methods to extend their solution to any other point
in space.
\begin{quote}
\textbf{Homework Problem: }
Find the magnetic field on the $z$-axis produced by a circular
current loop with radius $R$
lying in the $xy$-plane, centered at the origin,
carrying current $I$.
Do this analytically, then
show how to use Haskell to find the magnetic field anywhere.
\end{quote}

\section{Conclusion}

We've shown a number of features that functional programming languages have,
especially types, higher-order functions, and referential transparency,
that make them particularly appropriate for expressing the
elegant ideas of electromagnetic theory.
We've seen how scalar fields and vector fields can be encoded as types,
and how these types make central ideas of EM theory clear.
Aspects of vector calculus, like line and surface integrals,
can be used to express the Maxwell equations.
A Haskell version of the Biot-Savart law is arguably more understandable
than its traditional mathematical form.
Additional examples of electromagnetic theory in Haskell can be found
in \cite{walck2023learn}.
My anecdotal evidence that this is an effective way to teach the subject
comes from students this year performing better on the \emph{analytical}
portions of exams than in prior years.  Although the number of students
(10) is small, I take this to mean that the process of programming
in Haskell helped students get a better handle on the basic ideas
I'm trying to teach.
The author anticipates that dependently-typed functional languages
could have even more to offer to physics and EM theory.
For those who want to express the ideas of electromagnetic theory
in beautiful code, functional languages are hard to beat.

\nocite{*}
\bibliographystyle{eptcs}
\bibliography{em1}

\begin{thebibliography}{1}
\providecommand{\bibitemdeclare}[2]{}
\providecommand{\surnamestart}{}
\providecommand{\surnameend}{}
\providecommand{\urlprefix}{Available at }
\providecommand{\url}[1]{\texttt{#1}}
\providecommand{\href}[2]{\texttt{#2}}
\providecommand{\urlalt}[2]{\href{#1}{#2}}
\providecommand{\doi}[1]{doi:\urlalt{https://doi.org/#1}{#1}}
\providecommand{\eprint}[1]{arXiv:\urlalt{https://arxiv.org/abs/#1}{#1}}
\providecommand{\bibinfo}[2]{#2}

\bibitemdeclare{book}{sicp}
\bibitem{sicp}
\bibinfo{author}{Harold \surnamestart Abelson\surnameend},
  \bibinfo{author}{Gerald~Jay \surnamestart Sussman\surnameend} \&
  \bibinfo{author}{Julie \surnamestart Sussman\surnameend}
  (\bibinfo{year}{1996}): \emph{\bibinfo{title}{Structure and Interpretation of
  Computer Programs}}, \bibinfo{edition}{second} edition.
\newblock \bibinfo{publisher}{MIT Press}.

\bibitemdeclare{book}{griffiths2017introduction}
\bibitem{griffiths2017introduction}
\bibinfo{author}{David~J. \surnamestart Griffiths\surnameend}
  (\bibinfo{year}{2017}): \emph{\bibinfo{title}{Introduction to
  Electrodynamics}}.
\newblock \bibinfo{publisher}{Cambridge University Press},
  \doi{10.1017/9781108333511}.

\bibitemdeclare{book}{papert}
\bibitem{papert}
\bibinfo{author}{Seymour~A. \surnamestart Papert\surnameend}
  (\bibinfo{year}{1993}): \emph{\bibinfo{title}{Mindstorms: Children,
  Computers, And Powerful Ideas}}, \bibinfo{edition}{2} edition.
\newblock \bibinfo{publisher}{Basic Books}.

\bibitemdeclare{book}{sicm}
\bibitem{sicm}
\bibinfo{author}{Gerald~Jay \surnamestart Sussman\surnameend} \&
  \bibinfo{author}{Jack \surnamestart Wisdom\surnameend}
  (\bibinfo{year}{2001}): \emph{\bibinfo{title}{Structure and Interpretation of
  Classical Mechanics}}.
\newblock \bibinfo{publisher}{The MIT Press}.

\bibitemdeclare{book}{sussmanFDG}
\bibitem{sussmanFDG}
\bibinfo{author}{Gerald~Jay \surnamestart Sussman\surnameend} \&
  \bibinfo{author}{Jack \surnamestart Wisdom\surnameend}
  (\bibinfo{year}{2013}): \emph{\bibinfo{title}{Functional Differential
  Geometry}}.
\newblock \bibinfo{publisher}{The MIT Press}.

\bibitemdeclare{inproceedings}{walck2014}
\bibitem{walck2014}
\bibinfo{author}{Scott~N. \surnamestart Walck\surnameend}
  (\bibinfo{year}{2014}): \emph{\bibinfo{title}{Learn Physics by Programming in
  Haskell}}.
\newblock In \bibinfo{editor}{James \surnamestart Caldwell\surnameend},
  \bibinfo{editor}{Philip \surnamestart H{\"o}lzenspies\surnameend} \&
  \bibinfo{editor}{Peter \surnamestart Achten\surnameend}, editors: {\slshape
  \bibinfo{booktitle}{{\rm Proceedings 3rd International Workshop on} Trends in
  Functional Programming in Education, {\rm Soesterberg, The Netherlands, 25th
  May 2014}}}, {\slshape \bibinfo{series}{Electronic Proceedings in Theoretical
  Computer Science}} \bibinfo{volume}{170}, \bibinfo{publisher}{Open Publishing
  Association}, pp. \bibinfo{pages}{67--77}, \doi{10.4204/EPTCS.170.5}.

\bibitemdeclare{inproceedings}{walck2016}
\bibitem{walck2016}
\bibinfo{author}{Scott~N. \surnamestart Walck\surnameend}
  (\bibinfo{year}{2016}): \emph{\bibinfo{title}{Learn Quantum Mechanics with
  Haskell}}.
\newblock In \bibinfo{editor}{Johan \surnamestart Jeuring\surnameend} \&
  \bibinfo{editor}{Jay \surnamestart McCarthy\surnameend}, editors: {\slshape
  \bibinfo{booktitle}{{\rm Proceedings of the 4th and 5th International
  Workshop on} Trends in Functional Programming in Education, {\rm
  Sophia-Antipolis, France and University of Maryland College Park, USA, 2nd
  June 2015 and 7th June 2016}}}, {\slshape \bibinfo{series}{Electronic
  Proceedings in Theoretical Computer Science}} \bibinfo{volume}{230},
  \bibinfo{publisher}{Open Publishing Association}, pp.
  \bibinfo{pages}{31--46}, \doi{10.4204/EPTCS.230.3}.

\bibitemdeclare{book}{walck2023learn}
\bibitem{walck2023learn}
\bibinfo{author}{Scott~N. \surnamestart Walck\surnameend}
  (\bibinfo{year}{2023}): \emph{\bibinfo{title}{Learn Physics with Functional
  Programming: A Hands-on Guide to Exploring Physics with Haskell}}.
\newblock \bibinfo{publisher}{No Starch Press}.

\bibitemdeclare{misc}{diagrams}
\bibitem{diagrams}
\bibinfo{author}{Brent \surnamestart Yorgey\surnameend}
  (\bibinfo{year}{2011--2016}): \emph{\bibinfo{title}{The diagrams package}}.
\newblock
  \bibinfo{howpublished}{\url{https://hackage.haskell.org/package/diagrams}}.

\end{thebibliography}

\end{document}